\RequirePackage[2020-02-02]{latexrelease}
\documentclass[prl,reprint,superscriptaddress,preprintnumbers]{revtex4-1}
\pdfoutput=1
\usepackage{amsfonts, amssymb, amsmath}
\usepackage{mathrsfs}
\usepackage{hyperref}
\usepackage{tikz}
\usepackage[table]{}
\usepackage{empheq}

\usepackage{tensor}

\usepackage{color}
\definecolor{dark-gray}{gray}{0.20}
\definecolor{gray}{gray}{0.30}
\definecolor{light-gray}{gray}{0.80}
\definecolor{dark-red}{rgb}{0.7,0,0}
\definecolor{dark-green}{rgb}{0.1,0.4,0}
\definecolor{dark-blue}{rgb}{0.3,0.3,0.7}
\definecolor{light-blue}{rgb}{0.8,0.8,1}
\definecolor{blue}{rgb}{0,0,1}
\definecolor{red}{rgb}{1,0,0}
\definecolor{green}{rgb}{0,1,0}

\usepackage{hyperref}
\hypersetup{
	colorlinks=true,
	linkcolor=dark-blue,
	citecolor=dark-red,
	urlcolor=dark-blue,
	linktoc=page
}

\def\SU{{\rm SU}}

\def\ri{{\rm i}}

\newcommand{\be}{\begin{equation}}
\newcommand{\ee}{\end{equation}}
\newcommand{\bea}{\begin{eqnarray}}
\newcommand{\eea}{\end{eqnarray}}

\newcommand{\U}{\text{U}}

\newcommand\nn{\nonumber}
\newcommand\tx{\tilde x}
\newcommand\tu{\tilde u}
\newcommand\mn{\mathfrak n}

\newcommand\tB{\tilde B}

\newcommand\tv{\tilde v}

\usepackage{tikz}

\usepackage{xcolor}

\newcommand\fft[2]{\frac{#1}{#2}}

\def\SU{{\rm SU}}
\def\U{{\rm U}}

\begin{document}

\title{Large $N$ Topologically Twisted Indices, Holography, and Black Holes}

\date{\today}

\author{Nikolay Bobev}

\affiliation{Instituut voor Theoretische Fysica, KU Leuven, Celestijnenlaan 200D, B-3001 Leuven, Belgium}

\author{Junho Hong}

\affiliation{Instituut voor Theoretische Fysica, KU Leuven, Celestijnenlaan 200D, B-3001 Leuven, Belgium}

\author{Valentin Reys}

\affiliation{Instituut voor Theoretische Fysica, KU Leuven, Celestijnenlaan 200D, B-3001 Leuven, Belgium}

\begin{abstract}
\noindent We present a simple closed form expression for the topologically twisted index of the ABJM theory as a function of the magnetic fluxes and complex chemical potentials valid at fixed $k$ and to all orders in the $1/N$ expansion. This in turn leads to analytic expressions for the topologically twisted index at fixed genus in the 't~Hooft limit to all orders in the $1/\sqrt{\lambda}$ expansion. These results have important implications for holography and the microscopic entropy counting of supersymmetric static AdS$_4$ black holes. Generalizations to other SCFTs arising from M2-branes are also briefly discussed. 
\end{abstract}


\maketitle
%
\section{Introduction}\label{sec:intro}
%

Supersymmetric localization is a powerful tool for exact calculations of physical observables in strongly interacting supersymmetric QFTs. In the context of AdS/CFT it has been extensively utilized, see \cite{Pestun:2016zxk,Zaffaroni:2019dhb} for reviews, to provide new precision tests of the duality, gain new insights into black hole physics, and establish a new quantitative vantage point into the dynamics of string and M-theory. A notable class of examples that provide a fertile ground for explicit calculations are the 3d $\mathcal{N}=2$ SCFTs arising from $N$ M2-branes in M-theory. Employing supersymmetric localization, the $S^3$ partition function of such SCFTs reduces to a matrix model which in favorable circumstances can be solved in the large $N$ limit, see \cite{Fuji:2011km,Marino:2011eh,Mezei:2013gqa,Hatsuda:2012dt,Nosaka:2015iiw} and \cite{Marino:2016new} for a review. These results can then be used to explicitly test the AdS/CFT duality, to constrain the form of the higher-derivative corrections to the effective action of string and M-theory \cite{Chester:2018aca,Binder:2018yvd}, or to study four-derivative terms in 4d $\mathcal{N}=2$ gauged supergravity \cite{Bobev:2020pjk,Bobev:2021oku}. 

Our goal here is to derive similar results for the partition function of such models on $S^1\times \Sigma_{\mathfrak{g}}$ with $\Sigma_{\mathfrak{g}}$ a compact Riemann surface of genus $\mathfrak{g}$. This supersymmetric observable, also known as the topologically twisted index, was introduced in \cite{Benini:2015noa,Benini:2015eyy} where it was shown that it can be written as a finite-dimensional integral using supersymmetric localization. In the holographic context the large $N$ limit of the twisted index can be identified with the regularized on-shell action of certain supersymmetric Euclidean supergravity solutions, see \cite{Bobev:2020pjk}. In some instances these supergravity backgrounds admit analytic continuation to Lorentzian signature where they correspond to regular static dyonic BPS black holes in AdS$_4$. In these cases the leading order $N^{\frac{3}{2}}$ term in the topologically twisted index accounts for the Bekenstein-Hawking entropy of the black hole. The subleading $N^{\frac{1}{2}}$ and $\log N$ corrections to the topologically twisted index of the ABJM theory and the corresponding black hole entropy were discussed in \cite{Liu:2017vll,Liu:2017vbl,Bobev:2020pjk,Bobev:2021oku}. In this work we extend these leading and subleading order results to all orders in the $1/N$ expansion.

Our main focus here is on the topologically twisted index of the ABJM theory \cite{Aharony:2008ug}. This $\U(N) \times \U(N)$ Chern-Simons-matter theory describes $N$ M2-branes on $\mathbb{C}^4/\mathbb{Z}_k$ and preserves $\mathcal{N}=6$ supersymmetry for $k>2$. The Chern-Simons level $k$ plays the role of the inverse gauge coupling squared and one can perform a 't~Hooft expansion of observables in the theory with $\lambda = N/k$ being the 't Hooft coupling. In the limit of fixed $k$ and large $N$, the  ABJM theory is dual to the M-theory background AdS$_4 \times S^7/\mathbb{Z}_k$ where the sphere orbifold is freely acting. In this M-theory limit, the AdS radius $L$ is related to the gauge theory parameters as $(L/\ell_\mathrm{P})^6 \sim k\,N$ where $\ell_\mathrm{P}$ is the 11d Planck length, see e.g. \cite{Marino:2016new}. Since $S^7$ can be seen as a circle fibration over $\mathbb{CP}^3$, one can perform a reduction from M-theory to type IIA string theory and obtain a string background of the form AdS$_4 \times \mathbb{CP}^3$. To do perturbation theory in this duality frame the fibration circle should be small, which is achieved for large $k$. The string coupling constant is given by $k\,g_\mathrm{st} = L/\ell_\mathrm{s}\sim\lambda^{1/4}$ where $\ell_\mathrm{s}$ is the string length. The perturbative regime of type IIA string theory, characterized by large $k$ and small $g_\mathrm{st}$, can therefore be explored in the limit of fixed $\lambda$ and large $N$.

Our results for the topologically twisted index amount to closed form expressions for these two types of perturbative series and thus encode a wealth of information for the dual string and M-theory. It is expected that these resummed perturbative results receive non-perturbative corrections. While we are not able to compute these corrections explicitly, based on the results for the $S^3$ partition function we expect that in the limit of fixed $k$ and large $N$ there are $e^{-\sqrt{N k}}$ and $e^{-\sqrt{N /k}}$ corrections. Our analysis confirms these expectations.

%
\section{The sphere partition function}
\label{sec:S3}
%

To illustrate the type of exact expressions arising from supersymmetric localization we focus on the partition function of the ABJM theory on $S^3$ which serves as a blueprint for our analysis. As shown in \cite{Fuji:2011km,Marino:2011eh}, the perturbative part of the large~$N$ ABJM partition function on $S^3$ at fixed $k$ is given in terms of an Airy function and reads
\begin{equation}
\label{eq:Z}
	Z_{S^3} = C_k^{-\frac{1}{3}}\,e^{\mathcal{A}(k)}\,\text{Ai}\bigl[C_k^{-\frac{1}{3}}\bigl(N - B_k\bigr)\bigr] \, ,
\end{equation}
where
\begin{equation}
\label{eq:C-B}
	C_k = \frac{2}{\pi^2k} \, , \quad \text{and} \quad B_k = \frac{k}{24} + \frac{1}{3k}\, .
\end{equation}
The function $\mathcal{A}(k)$ entering \eqref{eq:Z} is given by
\begin{align}
\label{eq:A-ABJM}
	\mathcal{A}(k) =&\; \frac{2\zeta(3)}{\pi^2 k}\Bigl(1 - \frac{k^3}{16}\Bigr) + \frac{k^2}{\pi^2}\int_0^{\infty}\frac{x\log\bigl(1 - e^{-2x}\bigr)}{e^{kx} - 1}\,dx \nonumber \\[1mm]
	=& -\frac{\zeta(3)}{8\pi^2}k^2 + \frac12\log(2) + 2\zeta'(-1) + \frac16\log\Bigl(\frac{\pi}{2k}\Bigr) \nonumber \\
	& + \sum_{n\geq2}\Bigl(\frac{2\pi}{k}\Bigr)^{2n-2}\frac{(-1)^{n+1}4^{n-1}|B_{2n}B_{2n-2}|}{n(2n-2)(2n-2)!} \, ,
\end{align}
where the second equality shows the large $k$ expansion in terms of Bernoulli numbers $B_n$. With this data, we can expand the $S^3$ free energy $F_{S^3} = -\log Z_{S^3}$ in the M-theory limit ($k$ fixed and~$N\rightarrow\infty$) using the known asymptotic expansion of the Airy function to any desired order. The result to order $\mathcal{O}(N^{-\frac52})$ reads
\begin{align}
\label{eq:F-S3}
	F_{S^3} =&\; \frac{\pi\sqrt{2k}}{3}N^{\frac{3}{2}} - \frac{(8+k^2)\pi}{24\sqrt{2k}}N^{\frac{1}{2}} + \frac14\log N - \mathcal{A}(k) \nonumber \\ 
	& - \frac14\log\frac{k}{32} + \frac{(8+k^2)^2\pi}{2304\sqrt{2k}\,k}\,\frac{1}{N^{\frac{1}{2}}} - \frac{8+k^2}{96k}\,\frac{1}{N} \\
	& + \frac{(8+k^2)^3\pi^2 + 69120 k^2}{331776\pi\sqrt{2k}\,k^2}\,\frac{1}{N^{\frac{3}{2}}} - \frac{(8+k^2)^2}{4608k^2}\,\frac{1}{N^{2}} \, . \nonumber
\end{align}
We can also expand the free energy in the IIA limit ($\lambda$ fixed and $N \rightarrow \infty$) and write 
\begin{equation}
\label{eq:IIA-expand}
	F_{S^3} = -\sum_{{\tt g} \geq 0}\,(2\pi \ri\lambda)^{2{\tt g} - 2}F_{{\tt g}}(\lambda)\,N^{2 - 2{\tt g}} \, .
\end{equation}
In Type IIA string theory, this corresponds to a genus expansion of the string free energy, and the coefficients $F_{{\tt g}}$ are the genus-${\tt g}$ free energies at fixed 't~Hooft coupling $\lambda$. Rewriting the M-theory expansion \eqref{eq:F-S3} as a IIA expansion, we find the following free energies at low genera:
\begin{align}
\label{eq:Fg-S3}
	F_0(\lambda) =&\; \frac{4\pi^3\sqrt{2}}{3}\,\hat{\lambda}^{\frac{3}{2}} + \frac{\zeta(3)}{2} \, , \nonumber \\
	F_1(\lambda) =&\; \frac{\pi}{3\sqrt{2}}\,\hat{\lambda}^{\frac{1}{2}} - \frac14\log\hat{\lambda} + \frac16\log \lambda - \frac16\log N\nn\\
	&\; +\frac1{12}\log\frac{\pi^2}{32} + 2\zeta'(-1) - \frac12\log2 \, , \nonumber \\
	F_2(\lambda) =&\; \frac{5\,\hat{\lambda}^{-\frac{3}{2}}}{96\pi^3\sqrt{2}} - \frac{\hat{\lambda}^{-1}}{48\pi^2} + \frac{\hat{\lambda}^{-\frac{1}{2}}}{144\pi\sqrt{2}} - \frac{1}{360} \, , \\
	F_3(\lambda) =&\; \frac{5\,\hat{\lambda}^{-3}}{512\pi^6} - \frac{5\,\hat{\lambda}^{-\frac{5}{2}}}{768\pi^5\sqrt{2}} + \frac{\hat{\lambda}^{-2}}{1152\pi^4} \nonumber \\
	&\;- \frac{\hat{\lambda}^{-\frac{3}{2}}}{10368\pi^3\sqrt{2}} - \frac{1}{22680} \, , \nonumber
\end{align}
where we introduced the shifted 't~Hooft coupling $\hat{\lambda} = \lambda - \frac1{24}$ and included a $\log N$ term in $F_1$. We emphasize that \eqref{eq:Fg-S3} gives the genus-$\tt g$ free energies for all values of $\hat{\lambda} > 0$, up to non-perturbative corrections of order $\mathcal{O}(e^{-\sqrt{\lambda}})$. It is straightforward to extend these results to higher genera, and, interestingly, they can be successfully compared to the free energies obtained from studying the topological string on local $\mathbb{P}^1 \times \mathbb{P}^1$ \cite{Huang:2006si,Drukker:2010nc}. More details on this can be found in \cite{Drukker:2010nc} and in the upcoming work \cite{Bobev:2022eus}.\\

The $S^3$ partition function \eqref{eq:Z} can be deformed by turning on three real masses $m_{1,2,3}$ associated with the Cartan generators of the flavor symmetry group and by placing the theory on a squashed 3-sphere $S^3_b$ with $\U(1)\times \U(1)$ invariance. Here we conjecture that this deformed $Z_{S^3_b}(m_i)$ partition function is given by an Airy function as in \eqref{eq:Z}, with the following parameters:
\begin{equation}
	C_k = \frac{2}{\pi^2k}\frac{(b + b^{-1})^{-4}}{\prod_{a=1}^4 \Delta_a} \, , \quad B_k = \frac{k}{24} + \frac{\alpha(\Delta,b)}{k} \, ,\label{CB:general}
\end{equation}
where we have introduced the constrained quantities
\begin{equation}
\begin{split}
	\Delta_1 =&\; \frac12 - \ri\,\frac{m_1 + m_2 + m_3}{b+b^{-1}} \, , \; \Delta_2 = \frac12 - \ri\,\frac{m_1 - m_2 - m_3}{b+b^{-1}} \, , \\
	\Delta_3 =&\; \frac12 + \ri\,\frac{m_1 + m_2 - m_3}{b+b^{-1}} \, , \; \Delta_4 = \frac12 + \ri\,\frac{m_1 - m_2 + m_3}{b+b^{-1}} \, , \nonumber
\end{split}
\end{equation}
such that $\sum_a \Delta_a = 2$, and
\begin{equation}
	\alpha(\Delta,b) = -\frac1{12}\,\sum_{a=1}^4\Delta_a^{-1} + \frac{1 - \frac14\sum_a\Delta_a^2}{3(b+b^{-1})^2\prod_{a=1}^4 \Delta_a} \, .
\end{equation}
At present, we do not know the function that plays the role of $\mathcal{A}(k)$ in \eqref{eq:Z} for general $(m_i,b)$ but it is known in various limits. We have checked that our Airy conjecture for $Z_{S^3_b}(m_i)$ satisfies the relations put forward in \cite{Chester:2021gdw}, and reduces to known results in different limits \cite{Nosaka:2015iiw,Hatsuda:2016uqa}. More details on this will be presented in \cite{Bobev:2022eus} (A similar conjecture was independently proposed in \cite{Hristov:2022lcw}).

\section{The topologically twisted index}

The topologically twisted index for 3d $\mathcal{N}=2$ SCFTs on $S^1\times \Sigma_{\mathfrak{g}}$ was defined and studied in \cite{Benini:2015noa,Benini:2016hjo,Closset:2016arn}. It was shown in \cite{Benini:2015eyy,Benini:2016rke} that for generic parameters and for $\mathfrak{g}\neq 1$ the large $N$ limit of the ABJM index scales as $N^{\frac32}$. The $S^1 \times T^2$ index is more subtle and we will not discuss it here. From now on we take $\mathfrak{g}\neq 1$ and for brevity present only the final form of the index used in our calculations:
\begin{widetext}
	\begin{equation}
	\label{TTI:3}
		\begin{split}
			Z_{S^1\times  \Sigma_{\mathfrak{g}}}&=\prod_{a=1}^4y_a^{-\fft{N^2}{2}\mn_a}\sum_{\{x_i,\tx_j\}\in\text{BAE}}\left[\fft{1}{\det\mathbb B}\,\fft{\prod_{i=1}^Nx_i^N\tx_i^N\prod_{i\neq j}^N(1-\fft{x_i}{x_j})(1-\fft{\tx_i}{\tx_j})}{\prod_{i,j=1}^N\prod_{a=1}^2(\tx_j-x_iy_a)^{1-\frac{\mn_a}{1-\mathfrak{g}}}\prod_{a=3}^4(x_i-\tx_jy_a)^{1-\frac{\mn_a}{1-\mathfrak{g}}}}\right]^{1-\mathfrak{g}} \, .
		\end{split}
	\end{equation}
\end{widetext}
This expression is the result of applying supersymmetric localization to the path integral defining the ABJM twisted index and then evaluating the resulting finite-dimensional integral by residues employing a particular contour prescription. In \eqref{TTI:3}, the variables $x_i$ and $\tilde{x}_j$ are complex numbers that obey the so-called Bethe Ansatz Equations~(BAE),
\begin{equation}
\label{BA:operators}
	\begin{split}
		e^{\ri B_i}&\equiv x_i^k\prod_{j=1}^N\fft{(1-y_3\fft{\tx_j}{x_i})(1-y_4\fft{\tx_j}{x_i})}{(1-y_1^{-1}\fft{\tx_j}{x_i})(1-y_2^{-1}\fft{\tx_j}{x_i})}=1 \, , \\
		e^{\ri \tB_j}&\equiv \tx_j^k\prod_{i=1}^N\fft{(1-y_3\fft{\tx_j}{x_i})(1-y_4\fft{\tx_j}{x_i})}{(1-y_1^{-1}\fft{\tx_j}{x_i})(1-y_2^{-1}\fft{\tx_j}{x_i})}=1 \, ,
	\end{split}
\end{equation}
and the Jacobian matrix $\mathbb B$ is given by
\begin{equation}
	\mathbb B = \fft{\partial(e^{\ri B_1},\cdots,e^{\ri B_N},e^{\ri\tB_1},\cdots,e^{ \ri\tB_N})}{\partial(\log x_1,\cdots,\log x_N,\log\tx_1,\cdots,\log\tx_N)} \, .
\end{equation}
The chemical potentials $\Delta_a$, appearing in \eqref{TTI:3}~as~$y_a = e^{\ri\pi\Delta_a}$, and the magnetic charges $\mathfrak{n}_a$ for the $\U(1)^4$ Cartan subalgebra of the global symmetry group of the ABJM theory must satisfy the supersymmetry constraints
\begin{equation}
\label{eq:constraints}
	\sum_{a=1}^4 \Delta_a = 2 \, , \qquad \sum_{a=1}^4 \mathfrak{n}_a = 2(1 - \mathfrak{g}) \,.
\end{equation}
To evaluate the twisted index \eqref{TTI:3}, one needs to first solve the BAE and then take the sum over all BAE solutions in \eqref{TTI:3}. We now proceed to discuss a numerical approach to implement this procedure at large but finite $N$.

\section{The large $N$ limit}

Introducing the variables $u_i$ and $\tu_j$ through $x_i = e^{\ri u_i}$ and $\tx_j = e^{\ri\tu_j}$, an Ansatz for the BAE solutions in the large $N$ limit was constructed in \cite{Benini:2015eyy} and reads
\begin{equation}\label{eq:uidef}
	u_i = \ri\,N^{\frac12}\,t_i + v_i \, , \quad \tu_j = \ri\,N^{\frac12}\,t_j + \tv_j \, ,
\end{equation}
where $t_i$, $v_i$ and $\tv_i$ are real parameters that do not scale with $N$. We use the leading order BAE solution obtained from \eqref{eq:uidef} as an initial condition to find numerical BAE solutions for finite large values of $N$, using \texttt{FindRoot} in \textit{Mathematica} at \texttt{WorkingPrecision} $= 200$, see \cite{Liu:2017vll} for a similar approach. We assume that this BAE solution is the dominant contribution to the topologically twisted index in the large $N$ limit. For given fixed values of the parameters $k$, $\Delta_a$, and $\mn_a$ satisfying the constraints \eqref{eq:constraints}, we find numerical BAE solutions with values of $N$ in the range $[100,300]$ in steps of $10$. Substituting these large $N$ numerical solutions into \eqref{TTI:3}, we evaluate the twisted index for this particular $k$, $\Delta_a$, and $\mn_a$ and then use a \texttt{LinearModelFit} with respect to $N$ to determine the coefficients in the large $N$ expansion of the twisted index. We then repeat this process for numerous $k$, $\Delta_a$, and $\mn_a$ values to obtain the final expression of the twisted index for generic $k$, $\Delta_a$, and $\mn_a$ presented below. More details of this numerical analysis will be presented in a longer paper \cite{Bobev:2022eus}, and below we only present the final result. 

For the compact presentation of our results it proves useful to define the quantity
\begin{equation}\label{eq:Ndeltadef}
	\hat{N}_{\Delta} \equiv N-\fft{k}{24}+\fft{1}{12k}\,\sum_{a=1}^4 \Delta_a^{-1} \, ,
\end{equation}
in terms of which the logarithm of the perturbative part of the ABJM twisted index, $F_{S^1\times \Sigma_\mathfrak{g}} = -\log Z_{S^1\times \Sigma_\mathfrak{g}}$, takes the simple form:
\begin{align}
\label{TTI:exact}
	F_{S^1\times \Sigma_\mathfrak{g}} &\!= \fft{\pi\sqrt{2k\Delta_1\Delta_2\Delta_3\Delta_4}}{3}\sum_{a=1}^4\fft{\mn_a}{\Delta_a}\Bigl(\hat{N}_{\Delta}^\fft32 -\fft{\mathfrak c_a}{k}\hat{N}_{\Delta}^\fft12\Bigr) \nonumber \\
	&\;\;\;+\frac{1-\mathfrak{g}}{2}\log\hat{N}_{\Delta} - \hat f_0(k,\Delta,\mn)  \, ,
\end{align}
where $\mathfrak c_a$ are given by
\begin{equation}
	\mathfrak c_a=\fft{\prod_{b\neq a}(\Delta_a + \Delta_b)}{8\Delta_1\Delta_2\Delta_3\Delta_4}\,\sum_{b\neq a}\Delta_b \, , 
\end{equation}
and $\hat f_0$ is an undetermined function that does not depend on $N$. Note that \eqref{eq:Ndeltadef} agrees with the $b$-independent part of the shifted $N-B_k$ argument in the $S^3_b$ partition function, see \eqref{CB:general}. We also stress that \eqref{TTI:exact} is \emph{exact} up to non-perturbative corrections of order $\mathcal{O}(e^{-\sqrt{N k}})$ and $\mathcal{O}(e^{-\sqrt{N/k}})$ and therefore provides a compact resummation of the infinite series at large $N$ and fixed $k$. As stressed in \cite{Benini:2016rke} the twisted index is a meromorphic function of $y_a$ and thus the result above is valid if we substitute $\Delta_a$ with their complexification $\mathfrak{u}_a = \Delta_a+ \ri \beta \sigma_a$ where $\beta$ is the radius of the $S^1$ and $\sigma_a$ are the real mass parameters. While we have not provided an analytic derivation of the result in \eqref{TTI:exact} we have extensively checked it numerically with very high accuracy. In particular, we have confirmed numerically that it is accurate up to $\mathcal{O}(e^{-\sqrt{N}})$ corrections. Importantly, \eqref{TTI:exact} is consistent with previous results for the $N^\fft32$ leading order \cite{Benini:2015eyy} term, the first $N^\fft12$ correction \cite{Bobev:2020egg}, and the universal $\log N$ correction \cite{Liu:2017vll,Liu:2017vbl}. 

An important limit of the topologically twisted index is given by the so-called \textit{universal twist}, see \cite{Azzurli:2017kxo,Bobev:2017uzs}, when only the magnetic flux for the superconformal $\U(1)$ R-symmetry of the 3d SCFT is turned on. In our setup this amounts to setting $\Delta_a=\fft12$ and $\mn_a=\frac{1-\mathfrak{g}}{2}$. We then define $\hat{N} = N-\fft{k}{24}+\fft{2}{3k}$ to obtain 
\begin{equation}
\label{TTI:exact:minimal}
	\frac{F_{S^1\times \Sigma_\mathfrak{g}}}{1 - \mathfrak{g}} =\fft{\pi\sqrt{2k}}{3}\Bigl(\hat{N}^\fft32-\fft{3}{k}\hat{N}^\fft12\Bigr)+\fft12\log\hat{N} - \hat f_0(k) \, ,		
\end{equation}
where $\hat f_0(k)=\hat f_0(k,1/2,(1-\mathfrak{g})/2)/(1-\mathfrak{g})$. At present, we do not have a closed form expression for $\hat f_0(k)$ but we have been able to numerically obtain the first few terms in a large $k$ expansion:
\begin{equation}
\label{eq:A-S1S2}
	\begin{split}
		\hat f_0(k)=&\;-\fft{3\zeta(3)}{8\pi^2}k^2 + \fft76\log k + \mathfrak f_0 \\
		&\;+ \sum_{n=1}^{5}\Bigl(\frac{2\pi}{k}\Bigr)^{2n}\,\frac{\mathfrak f_{2n}}{3^{n+2}} + \mathcal{O}(k^{-12}) \, ,
	\end{split}
\end{equation}
with
\begin{equation}
	\{\mathfrak f_{2n}\}= \Bigl\{-\frac65,\,\frac{19}{70},\,-\frac{41}{175},\,\frac{279}{700},\,-\frac{964636}{875875}\Bigr\} \, 
\end{equation}
for $n=1,\cdots,5$ and $\mathfrak f_0 =-2.096848299$ (for brevity here and below we present only the first 10 digits of the numerical constants). For low values of $k$,  $\hat f_0(k)$ can be determined numerically with very good precision and we find the following numerical values: 
\begin{equation}
	\begin{split}
		\hat f_0(1) &= -3.045951311\,,  ~~~ \hat f_0(2)= -1.786597534\,, \\
		\hat f_0(3) &=-1.386373044\,, ~~~ \hat f_0(4) = -1.306589553\,.
	\end{split}
\end{equation}
It is important to stress that for any given choice of $(k,\Delta_a,\mn_a)$ the function $\hat f_0$ in \eqref{TTI:exact} can similarly be found to a high degree of numerical accuracy. 

Expanding \eqref{TTI:exact:minimal} at large $N$ and fixed $k$ allows for a direct comparison with the large $N$ expansion of the $S^3$ free energy \eqref{eq:F-S3}. As shown in \cite{Azzurli:2017kxo}, see also \cite{Hosseini:2016ume}, we recover the universal relation $F_{S^1\times \Sigma_\mathfrak{g}} = (1-\mathfrak{g})\,F_{S^3}$ at the leading $N^{\frac{3}{2}}$ order. However, there are differences in the coefficients of the subleading order terms. In particular, we recover the coefficients of the $N^{\frac{1}{2}}$ and $\log N$ terms found in  \cite{Liu:2017vll,Bobev:2020egg,Liu:2017vbl}. 

As in \eqref{eq:IIA-expand}, we can also convert the expression \eqref{TTI:exact:minimal} to a type IIA, or 't Hooft, expansion for the twisted index. This gives the following genus-${\tt g}$ free energies at low genera (we set $\mathfrak{g}=0$ for brevity, for the general result the left hand side has to be divided by $(1-\mathfrak{g})$):
\begin{align}
\label{eq:Fg-TTI}
	F_0(\lambda) =&\; \frac{4\pi^3\sqrt{2}}{3}\,\hat{\lambda}^{\frac{3}{2}} + \frac{3\zeta(3)}{2} \, , \nonumber \\
	F_1(\lambda) =&\; \frac{2\pi\sqrt{2}}{3}\,\hat{\lambda}^{\frac{1}{2}} - \frac12\log\hat{\lambda} - \frac23\log\lambda + \frac23\log N + \mathfrak{f}_0 \, , \nonumber \\
	F_2(\lambda) =&\; \frac{\hat{\lambda}^{-1}}{12\pi^2} - \frac{5\hat{\lambda}^{-\frac{1}{2}}}{36\sqrt{2}\pi} + \frac2{45} \, , \\	
	F_3(\lambda) =&\; \frac{\hat{\lambda}^{-2}}{144\pi^4} - \frac{\hat{\lambda}^{-\frac{3}{2}}}{162\sqrt{2}\pi^3} + \frac{19}{5670} \, . \nonumber
\end{align}
At large $\lambda$, the genus-0 free energy matches the one extracted from the $S^3$ free energy in \eqref{eq:Fg-S3}, while differences appear at finite 't~Hooft coupling and at higher genera. As in the $S^3$ case, one can also find explicit results for the ${\tt g}>3$ free energies. Note that the genus-0 and genus-1 results in \eqref{eq:Fg-TTI} agree with the numerical analysis in \cite{PandoZayas:2019hdb} in the large $\lambda$ limit, including the coefficient $-\frac{7}{6}$ of the $\log\lambda$ term. As in~\eqref{eq:Fg-S3}, the results for $F_{{\tt g}}$ are valid up to non-perturbative corrections of order $\mathcal{O}(e^{-\sqrt{\lambda}})$.

%
\section{Holography and black holes}
%

The simple result for the logarithm of the twisted index \eqref{TTI:exact} is striking and begs for an interpretation in the dual gravitational theory. When viewed as a Euclidean partition function the twisted index should be dual to the path integral of string/M-theory on an asymptotically locally AdS$_4$ Euclidean background. If one works in an ensemble with fixed magnetic charges $\mathfrak{n}_a$ and chemical potentials $\Delta_{a}$ the appropriate solutions in the two-derivative supergravity limit are the ``black saddles'' discussed in~\cite{Bobev:2020pjk}. The result in \eqref{TTI:exact} can then be viewed as a field theory prediction for the gravitational path integral on these backgrounds. The simplest example of such a supergravity background is the Euclidean Romans solution \cite{Romans:1991nq,Bobev:2020pjk} which should be dual to the twisted index with the universal twist in \eqref{TTI:exact:minimal}. For $\mathfrak{g}>1$ this solution admits a smooth analytic continuation into the extremal Lorentzian AdS$_4$ Reissner-Nordstr\"om black hole with a $\Sigma_{\mathfrak{g}}$ horizon and vanishing electric charges. As shown in  \cite{Benini:2015eyy,Azzurli:2017kxo}, \cite{Bobev:2020egg,Bobev:2021oku}, and \cite{Liu:2017vbl} the topologically twisted index with the universal twist correctly accounts for the leading $N^{\frac{3}{2}}$, subleading $N^{\frac{1}{2}}$, and sub-subleading $\log N$ terms in the entropy of this black hole. These results suggest that no dangerous cancellations due to the $(-1)^F$ factor in the index occur up to this order. If we assume that this remains true to all orders in the $1/N$ expansion, our result for the universal topologically twisted index in~\eqref{TTI:exact:minimal} amounts to a prediction for the black hole entropy to all orders in the large $N$ expansion. For general values of the magnetic fluxes and chemical potentials the twisted index in \eqref{TTI:exact} should also encode information about the physics of black hole microstates. Indeed, as shown in \cite{Benini:2015eyy,Benini:2016rke,Bobev:2018uxk}, after an appropriate Legendre transform and $\mathcal{I}$-extremization the $N^{\frac{3}{2}}$ term in the twisted index accounts for the Bekenstein-Hawking entropy of the dyonic asymptotically locally AdS$_4$ black holes constructed in \cite{Gauntlett:2001qs,Cacciatori:2009iz}. Extending this procedure to all orders in the large $N$ expansion is not trivial and will be discussed further in \cite{Bobev:2022eus}.

Another framework in which our result for the topologically twisted index \eqref{TTI:exact} can be applied to black hole physics is in the computation of the Quantum Entropy Function \cite{Sen:2008vm} for AdS black holes by means of supergravity localization \cite{Hristov:2018lod}. In particular, the structure of the result in \eqref{TTI:exact} can be used to constrain the relevant 4d supergravity bulk quantities such as the prepotential specifying the theory and the dimensionless ratio $L^2/G_N$ between the AdS$_4$ scale and the Newton constant. We will report on this in \cite{Bobev:2022eus}.

%
\section{Discussion}\label{sec:conclusion}
%

The all order compact expression for the large $N$ topologically twisted index of the ABJM theory in \eqref{TTI:exact} can be generalized to other holographic SCFTs as will be shown in detail in \cite{Bobev:2023lkx}. A particularly simple example is the mABJM 3d $\mathcal{N}=2$ SCFT, see \cite{Bobev:2018uxk}, for which the large $N$ twisted index can be found by setting  $\Delta_1=1$ and $\mathfrak{n}_1= (1 - \mathfrak{g})$ in \eqref{TTI:exact} while still imposing the constraints in \eqref{eq:constraints}. Another holographic theory for which the large $N$ index can be computed exactly is the 3d $\mathcal{N}=4$ $\SU(N)$ SYM theory coupled to one adjoint and $N_f$ fundamental hyper multiplets. The perturbative part of the topologically twisted index for this model in the large $N$, fixed $N_f$ limit can be written compactly in terms of the quantity $\tilde{N} = N+\fft{7N_f}{24}+\fft{1}{3N_f}$ for the universal twist as
\begin{equation}
\label{TTI:exact:ADHM}
	\begin{split}
	\frac{F_{S^1\times \Sigma_\mathfrak{g}}}{1 - \mathfrak{g}} =&\; \fft{\pi\sqrt{2}}{3}N_f^\fft12\Bigl(\tilde{N}^\fft32-\Bigl(\frac{N_f}{2}+\fft{5}{2N_f}\Bigr)\tilde{N}^\fft12\Bigr)\\
	&\;+\fft12\log\tilde{N}-\tilde{f}_0(N_f)\,.		
	\end{split}
\end{equation}
The leading $N^{\frac{3}{2}}$ term and the first subleading $N^{\frac{1}{2}}$ correction in this expression agree with the results in \cite{Hosseini:2016ume} and \cite{Bobev:2020egg}, respectively. The function $\tilde{f}_0(N_f)$ plays a similar role to $\hat{f}_0$ in \eqref{TTI:exact} and can be determined numerically with high precision. 

While we have confirmed the validity of \eqref{TTI:exact} with extensive numerical checks it will be most interesting to have an analytic derivation of this formula. Amongst other things, this will shed light on the precise analytic form of the function $\hat{f}_0$ as well as the details of the non-perturbative $\mathcal{O}(e^{-\sqrt{N k}})$ and $\mathcal{O}(e^{-\sqrt{N /k}})$ corrections. It is also important to understand whether there are other solutions to the BAE that contribute to the large $N$ limit of the twisted index and to uncover their dual gravitational interpretation. It is natural to wonder whether the simple compact form of the large $N$ twisted index and $S^3$ partition function of the ABJM theory point towards similar simple expressions for the partition function of the theory on other compact Euclidean manifolds, see \cite{Closset:2019hyt} for a review. It will therefore be interesting to revisit the supersymmetric localization calculations of these observables in the large $N$ limit and understand for which of them one can find closed form expressions.

Our results have important implications for holography and black hole physics. It will be interesting to understand whether the large $N$ twisted index in \eqref{TTI:exact} can be derived in supergravity or string/M-theory. A possible way to achieve this is to employ supersymmetric localization in supergravity. This was studied in \cite{Dabholkar:2014wpa} for the $S^3$ partition function of the ABJM theory and it will be interesting to perform a similar supergravity localization analysis for the supersymmetric asymptotically AdS$_4$ solutions dual to the topologically twisted index. The recent results in \cite{Hristov:2018lod,Hristov:2019xku,Hristov:2021zai} should serve as a steppingstone towards this analysis.

Alternatively, instead of reproducing our result in \eqref{TTI:exact} by an independent calculation in the gravitational side of AdS/CFT, it could be used to learn about the structure of higher-derivative corrections in M-theory. This approach has been utilized in \cite{Chester:2018aca,Binder:2018yvd} to relate the large $N$ results for the $S^3$ partition function of the ABJM theory to corrections to 11d supergravity. Studying this in more detail, and using the results in \cite{Bergman:2009zh}, should also shed light on the M-theory origin of the shift in $N$ in \eqref{eq:Ndeltadef} naturally suggested by our results. 

Given the relation between the $S^3$ partition function of the ABJM theory and topological strings \cite{Drukker:2010nc} together with the fact that the topologically twisted index is related to the $A$-model topological field theory on $\Sigma_{\mathfrak{g}}$ it is tempting to speculate that the ABJM topologically twisted index is also related to topological strings. It will be most interesting to uncover such a relation which will have intriguing implications for holography and black holes and may point to a generalization of the OSV conjecture \cite{Ooguri:2004zv} to asymptotically AdS$_4$ black holes.

%
%
\section*{Acknowledgements}
\noindent We are grateful to A.~M.~Charles, K.~Hristov, and Y.~Xin for useful discussions. We are supported in part by an Odysseus grant G0F9516N from the FWO and by the KU Leuven C1 grant ZKD1118 C16/16/005.

\bibliography{TTI-holo}

\end{document}